\documentclass[showpacs,prl,aps,twocolumn]{revtex4-1}

\pdfoutput = 1

\usepackage{amsmath,amsfonts,amssymb,bm,graphicx,hyperref,color,units}

\newcommand{\bra}[1]{\langle #1|}
\newcommand{\ket}[1]{|#1\rangle}
\newcommand{\Tr} {{\text T}{\text r}}

\begin{document}
\title{Fluctuation Theorems and the Generalised
Gibbs Ensemble in Integrable Systems}
\author{James M. Hickey}
\author{Sam Genway}

\affiliation{School of Physics and Astronomy, University of Nottingham,
Nottingham, NG7 2RD, United Kingdom}

\date{\today}

\begin{abstract}
We derive fluctuation relations for a many-body quantum system prepared in a
Generalised Gibbs Ensemble subject to a general nonequilibrium protocol.  By considering isolated integrable systems, we find generalisations to the Tasaki-Crooks and Jarzynski relations. 
Our approach is illustrated by studying the one-dimensional quantum Ising model subject to a sudden change in the transverse field, where we find that the statistics of the work
done and irreversible entropy show signatures of quantum criticality.  We
discuss these fluctuation relations in the context of thermalisation.

\end{abstract}

\maketitle
\emph{Introduction.}  The last decade has seen an explosion of interest, both experimental and
theoretical, in the exploration of out-of-equilibrium quantum systems.  A particular motivation has been the advance in ultracold-atom experiments which allow closed quantum systems to be realised, manipulated and probed~\cite{Bloch2002,Weiss2006,Cheneau2012,Gring2012}.  Driven by
this progress, theoretical investigations have revealed a number of interesting
phenomena involving quantum nonequilibrium dynamics, fluctuations and
thermalisation~\cite{Goldstein2006,Popescu2006,Polkovnikov2011,Marcuzzi2012, Lutz2013, Strasberg2013, Hickey2013,Heyl2012A,Gambassi2012}.  
In particular, much interest has arisen in the nonequilibrium fluctuations of a system when parametrically changing the Hamiltonian over time.   
When applying such nonequilibrium protocols to systems prepared initially in a thermal state, these fluctuations have been shown to obey Tasaki-Crooks and Jarzynski relations~\cite{Jarzynski2011,Jarzynski1997,Crooks1999,Tasaki2000,Mukamel2003}.

A much-studied nonequilibrium protocol is the global
quantum quench, whereby the system is prepared in a ground state $\ket{0}$ of
a specified Hamiltonian $H(\lambda_0)$ which depends parametrically on an experimentally-tunable global parameter $\lambda_0$.  This parameter is changed abruptly such that
$\lambda_0\rightarrow\lambda_t$, and the state $\ket{0}$ is allowed to evolve
coherently under the Hamiltonian: $\ket{\psi_t} = {e}^{-i
H(\lambda_t)t}\ket{0}$.  It has been shown that after a long time $t$, small
subsystems within the closed system can be described by statistical ensembles. 
Generically, the reduced density matrices of subsystems evolve towards thermal
states~\cite{Tasaki1998, Mahler2005, Linden2008, Reimann2010, *Reimann2008,
Genway2010,*Genway2012,*Genway2013, Cramer2011}.  However, for integrable
systems where there are additional local conserved quantities, subsystems are known to approach the so-called Generalised Gibbs Ensemble (GGE)~\cite{Rigol2007,Calabrese2011,*Calabrese2012,*Calabrese2012b,Essler2012, Fagotti2013,Caux2012,Deutsch1991,Srednicki1994}.  The GGE is specified by an extensive number of conserved charges $I_k$ on the global system and associated Lagrange multipliers $\mu_k$.

In this work, we consider one of the GGE subsystems of an integrable system in isolation, and ask about fluctuation phemonena.  By studying isolated integrable systems, we derive generalisations to the Tasaki-Crooks and Jarzynski relations for systems initially in a GGE.  We illustrate our results by applying them to the transverse-field Ising model (TFIM)~\cite{Sachdev2011} and calculate
generalised work and joint probability distributions, and the irreversible entropy associated with quench protocols.
Finally we discuss the implications of our results on quantum quenches and prethermalisation~\cite{Eckstein2008,*Kollar2011,Gring2012,Matteo2013}.

\emph{Generalised work and irreversible entropy.}  
We will study systems with Hamiltonian $H(\lambda_0)$ initially in states described by the GGE
\begin{equation}
\rho_{GGE}(\lambda_0) = \frac{e^{-\sum_{k}\mu_k I_k}}{Z}.
\end{equation}
where $I_k$ are conserved quantities with $[H(\lambda_0),I_k]=[I_k,I_{k'}]=0 \, \forall\, k,k'$, $\mu_k$ are associated Lagrange multipliers and the normalisation is defined by $Z = \Tr(e^{-\sum_k \mu_k I_k})$.  
%
 For times $t \ge 0$ up to the time $t=\tau$, work is done on the system under a protocol where $\lambda_0 \to \lambda_\tau$.  The final
Hamiltonian $H(\lambda_\tau)$ has a set of conserved quantities we will denote $J_k$.  We consider a continuous class of operators $I_k(\lambda)$ such that $I_k(\lambda_0) = I_k$ and $I_k(\lambda_\tau) = J_k$, in essence each quantity $J_k$ of $H(\lambda_\tau)$ is the continuation of the conserved quantity $I_k$ of $H(\lambda_0)$ with the same $k$.   

  We will adopt a two-point measurement
scheme to define a generalised work done~\cite{Esposito2009,Schon2007}.
We first consider performing a measurement on the initial set of conserved quantities $\{I_{k}\}$ which results in 
eigenvalues $\{\epsilon^k_0\}$;  this collapses the state of the system into an eigenstate of the set $\{I_{k}\}$.
We then evolve the system under unitary dynamics $U(\tau,0)$ describing the
 protocol $\lambda_0\rightarrow \lambda_\tau$ before measuring the eigenvalues $\{\epsilon^{k}_{\tau}\}$ associated with the final set of conserved quantities $\{J_{k}\}$, at time
$\tau$. We will use two
complete basis sets $\ket{m,\{\epsilon^{k}_{0}\}}$ and $\ket{n,\{\epsilon^{k}_{\tau}\}}$
where the indices $m~(n)$ distinguish between states with the same set of eigenvalues
$\{\epsilon^k_{0}\}~(\{\epsilon^k_{\tau}\})$ associated with $\{I_k\}$~($\{J_k\}$).

According to this measurement scheme,
the probability density,  \emph{i.e.} the probability for the generalised work done for the forward protocol, is given by
\begin{align}
\label{eq:eq1}
&P_{F}(W) =
\sum_{m,n,\{\epsilon^k_0\},\{\epsilon^k_\tau\}} \frac{{e}^{-\sum_k\mu_k
\epsilon^k_0}}{Z(\lambda_0)}  \\
&\times
\big|\bra{n,\{\epsilon^k_\tau\}}U(\tau,0)\ket{m,\{\epsilon^k_0\}}\big|^{2}  \delta\Big[W-\sum_k
(\tilde{\mu_k}\epsilon_{\tau}^{k}-\mu_k\epsilon^{k}_0)\Big], \nonumber
\end{align}
Multiplying both sides by
${e}^{-W}$ and rearranging we find
\begin{equation}
\label{eq:eq2}
\frac{P_F(W)}{P_{B}(-W)} ={e}^{W-\Delta F} ,
\end{equation}
where $\Delta F = \log \Tr({e}^{-\sum_k \mu_k
I_k})/\Tr({e}^{-\sum_k \tilde{\mu}_k J_{k}})$.  This is a
generalisation of the Tasaki-Crooks fluctuation relation and encodes the
relationship between the measurement distributions of the forward protocol, as
described above, and the \emph{backward} protocol.  The \emph{backward} work
distribution is obtained by preparing the system initially in the state
$\rho_{GGE}(\lambda_\tau) = \sum_{k}{e}^{- \tilde{\mu}_k J_{k}}/Z(\lambda_\tau)$ 
and evolving the system under the reverse protocol $\lambda_\tau \rightarrow \lambda_0$ defined by
$U^{\dagger}(\tau,0)$. Taking the Fourier transform of both sides of
Eq.~\eqref{eq:eq2} allows this fluctuation theorem to be expressed in terms of
characteristic functions~\cite{Talkner2007,Talkner2007b,Campisi2011,Campisi2008}:
\begin{equation}
\label{eq:eq3}
\frac{\chi_F(u,\tau)}{\chi_B(-u+i,\tau)} =
\frac{Z(\lambda_\tau)}{Z(\lambda_0)}.
\end{equation}
where
\begin{align}
\label{eq:Char}
\chi_F(u,\tau) &= \int
dW{e}^{iuW}P_F(W)  \\
&= \Tr(U^{\dagger}(\tau,0){e}^{iu\sum_k
\tilde{\mu}_k J_k}U(\tau,0){e}^{-iu\sum_k \mu_k I_k}\rho_{GGE}). \nonumber
\end{align}
Analytically continuing the characteristic function by setting $u
= i$ we obtain
\begin{equation}
\label{eq:eq4}
\chi_F(u=i) = \langle e^{-W}\rangle =
{e}^{-\Delta F} = \frac{Z(\lambda_\tau)}{Z(\lambda_0)},
\end{equation}
which is the extension of the Jarzynski equality~\cite{Jarzynski1997} to the GGE.  

The above relationships simplify if three requirements are met: the first is that the
Lagrange multipliers of the initial and final GGEs are equal, \emph{i.e.} $\mu_k =
\tilde{\mu_k}$, $\forall k$;   the second is that the
initial [$H(\lambda_0)$] and final [$H(\lambda_\tau)$] Hamiltonian be expressed as a tensor product of the Hilbert spaces spanned by the associated conserved quantities;  the final requirement is that conserved charges 
 $J_k$ and $I_k$ live in the same subspace $\forall k$
and the nonequilibrium protocol $U(\tau,0)$ does not connect the
different subspaces spanned by these conserved quantities.  Then, we can consider
measurements solely on individual $J_k$ and $I_k$ leading to the simpler analogs of
Eqs.~\eqref{eq:eq3} and~\eqref{eq:eq4}:
\begin{align}
\label{eq:eq5}
\frac{\chi_F(u,\tau)}{\chi_B(-u+i,\tau)} &=
\frac{Z_k(\lambda_\tau)}{Z_k(\lambda_0)}, \\
\chi_F(i) = \langle e^{-W}\rangle &=
{e}^{-\Delta F_{i}}, \nonumber
\end{align}
where $Z_k$ is the normalisation associated with the $k$th conserved charge and
$\Delta F_k = -\log{Z_k(\lambda_{\tau})/Z_k(\lambda_0)}$.  Later we will find that this simplification can arise when studying
a quench protocol, \emph{i.e.} $U(\tau,0) = 1$ as $\tau \rightarrow 0^{+}$.  We note that if the initial state is a thermal state, there is only one Lagrange multiplier, \emph{i.e.} $\mu_k = \tilde{\mu_k} = \beta$, and the conserved quantities of interest (the Hamiltonians) live within the same Hilbert space.  In this instance the above requirements hold necessarily and from Eq.~\eqref{eq:eq5} we obtain the standard Tasaki-Crooks and Jarzynski equalities~\cite{Dorner2012}.  

These fluctuation relations encompass all the information on the work done in an integrable model: regardless of how far the system is driven from equilibrium, they are determined by the equilibrium properties of the system in a GGE.  In the thermodynamic limit, the law of large numbers ensures that entropy of an
isolated never decreases under a thermodynamic process.  For a finite
size system the second law of thermodynamics must be extended and, in accordance
with the Jarzynski equality, the expectation value of the work done is greater
than or equal to the change in equilibrium free energy.  This extends from the canonical ensemble to GGEs, with Eq.~\eqref{eq:eq4} implying that
$\langle W \rangle \ge \Delta F$.  For
non-ideal processes, the difference between these two processes is the average
\emph{generalised irreversible work done},
\begin{equation}
\label{eq:eq6}
\langle W\rangle = \langle W^{irr}\rangle +\Delta F.
\end{equation}
As we are considering closed many-body systems, there can be no heat transfer and
therefore the change in entropy under our non-equilibrium protocol can only be
due to irreversible entropy production, which in our case is the irreversible
work done~\cite{Jarzynski1997,Esposito2009},
\begin{equation}
\label{eq:eq7}
\langle \Delta S^{irr} \rangle = \langle
W \rangle - \Delta F  = \langle W^{irr} \rangle.
\end{equation}

Until now we have studied a scheme where we measure \emph{all} of the conserved quantities $\{I_k\}$ and $\{J_k\}$ under changing $\lambda_0 \to \lambda_\tau$.  However, we can also consider the joint probability density for measuring one conserved quantities from each set namely $I_{k}$ and $J_{k'}$ at the beginning and end of the nonequilibrium protocol respectively.  The outcomes of these measurements are eigenvalues
$\epsilon^k_0$ and $\epsilon^{k'}_\tau$, where in general we consider $k'\neq
k$.  Taking two complete basis sets as above,  $\ket{m,\epsilon^k_0}$ and $\ket{n,\epsilon^{k'}_\tau}$, where $m~(n)$ now distinguishes between states with only one common eigenvalue $\epsilon^k_0~(\epsilon^{k'}_{\tau})$, we find a generic two-point joint probability density~\cite{Esposito2009,Schon2007}:
\begin{align}
\label{eq:eq8}
P^{TP}_F(W) =& \sum_{m,n,\epsilon^k_0,\epsilon^{k'}_\tau}
\bra{m,\epsilon^k_0}\rho_{GGE}(\lambda_0)\ket{m,\epsilon^k_0}
\\ & \times|\bra{n,\epsilon^{k'}_\tau}U(\tau,0)\ket{m,\epsilon^k_0}|^{2} 
\delta\big[W-(\epsilon_{\tau}^{k'}-\epsilon^{k}_0)\big]. \nonumber
\end{align}
Although similar to our previous considerations, here the ``generalised work done'' has no
special dependence on the chemical potentials and we consider it to be the most general ``work'' distribution which is directly related to the GGE form associated with integrable systems.


\emph{Example: transverse-field Ising model.} We now apply the theoretical
framework described previously to the concrete example of the one-dimensional transverse-field Ising model
(TFIM).  The TFIM is described by the Hamiltonian
\begin{equation}
H = -\sum_i \sigma^{z}_i\sigma^z_{i+1} -\lambda \sum_i \sigma^x_i.
\end{equation}
In the $N$-spin model with periodic boundary
conditions which we will study, the operators $\sigma^{x,z}_i$ are the Pauli operators at site $i$ and the
parameter $\lambda$ denotes the transverse field strength.  This is a
paradagmatic model of a quantum phase transition: in the thermodynamic limit the
ground state changes singularly at $\lambda = \pm 1$.  For $|\lambda|< 1$ the ground
state is ferromagnetic in nature while when $|\lambda|>1$ it is paramagnetic~\cite{Sachdev2011}.  This
Hamiltonian may be diagonalised exactly using textbook techniques employing Jordan-Wigner fermions and a Bogoluibov transformation with a set of Bogoluibov angles $\varphi_k$ which reduce the Hamiltonian to
\begin{align}
H &= \sum_k \epsilon_k(\lambda)(\gamma^{\dagger}_k\gamma_k - 1/2), \nonumber \\
  &= \sum_k H_k(\lambda_0).
\end{align}
Here, $\gamma_k~(\gamma^{\dagger}_k)$ is the fermionic quasiparticle
annihilation (creation) operator and $\epsilon_k(\lambda) =
2\sqrt{(\lambda-\cos k)^2 +\sin^2 k}$.  The wavevectors $k$ lie
in the range $[-\pi,\pi]$ and are given by $k = 2\pi n/N$ where $n = -N+1,-N+3,
\ldots, N-1$. (To be concrete, we focus on the case where $N$ is even.) 
The occupation numbers of each fermionic mode $\gamma^{\dagger}_k\gamma_k$ are the conserved quantities
in this model.  It has been shown that one can prepare a subsystem in a state $\rho_{GGE}(\lambda_0) \propto {e}^{-\sum_k \mu_k \gamma^{\dagger}_k \gamma_k }$ via a quantum quench from a state $\ket{\psi(\lambda_{t\ll0})}$ with the Lagrange multipliers $\mu_k$ fixed by the expectation values $\langle \gamma^{\dagger}_k\gamma_k\rangle =
\Tr[\rho_{GGE}(\lambda_0)\gamma^{\dagger}_k \gamma_k]$.  The eigenvalues of
these fermionic occupation operators $\gamma^{\dagger}_{k}\gamma_{k}$ are labelled $n_k = 0,1$.

Having prepared our system of interest in a GGE, we will consider a protocol implemented via instantaneous switch from $\lambda_0$ to $\lambda_\tau$, where the final Hamiltonian is
\begin{align}
H(\lambda_\tau) &= \sum_k
\epsilon_k(\lambda_\tau)(\tilde{\gamma}^{\dagger}_k\tilde{\gamma}_k -1/2), \nonumber \\
&= \sum_k H_k(\lambda_\tau).
\end{align}
Here we note that the pre- and post-quench fermionic operators are
necessarily different but related to each other through the difference in
Bogoliubov angles $\varphi_k$ which diagonalise the Hamiltonian.  Furthermore, the chemical potentials can be calculated for the initial and final GGEs~\cite{Polkovnikov2011,Rigol2007} by
$\mu_k (\tilde{\mu}_k) = -\log \tan^2 \alpha_k (-\log \tan^2 \tilde{\alpha}_k)
$ where $\alpha_k(\tilde{\alpha_k}) $ are the difference in Bogoliubov angles
which diagonalise $H(\lambda_{t\ll 0})$ and $H(\lambda_0) (H(\lambda_\tau))$.
In the TFIM, the fermionic occupations form a set of good quantum
numbers and the quench protocol implies $U(\tau,0) = 1$ as $\tau \rightarrow 0^{+}$.  The final fermionic operators after the quench given by
$J_k=\tilde{\gamma}^{\dagger}_k\tilde{\gamma}_k$ are linear combinations of
$\gamma^{\dagger}_{\pm k}$ and $\gamma_{\pm k}$. They therefore remain within the
$\pm k$ subspace of the original conserved quantities, and under the quench protocol satisfy the three requirements for the simplified Eq.~\eqref{eq:eq5} to hold. 

\begin{figure}
\includegraphics[scale = 1.0]{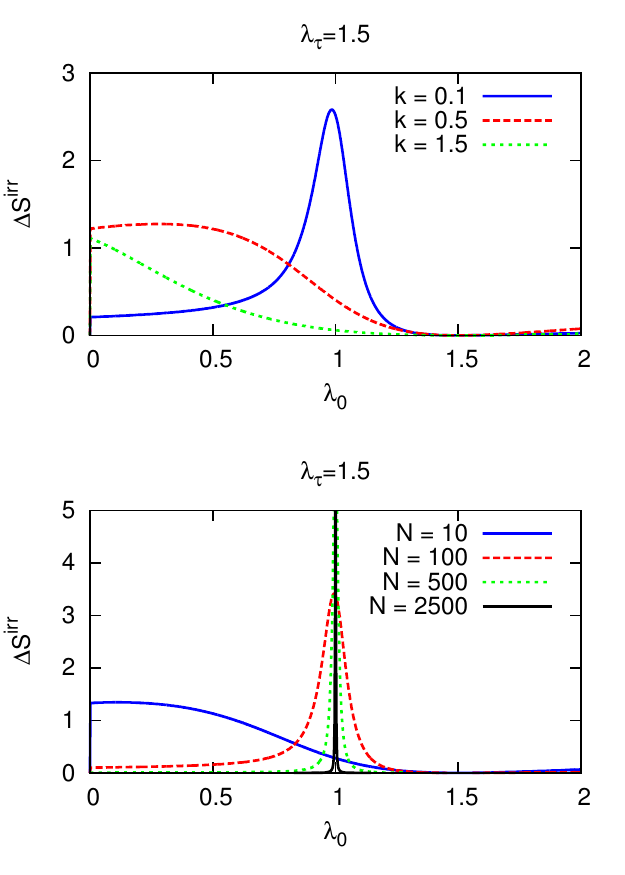}
\caption{(Color Online) Top: The irreversible entropy associated with the mode
$k$ is plotted as a function of the value $\lambda_0$.  This grows and sharpens
about the quantum critical point as $k\rightarrow 0$, marking the
irreversibility associated with criticality.  Bottom: The irreversible entropy
of the longest wavelength mode, $k = 2\pi/N$, grows and sharpens about the quantum critical point
with increasing system size.  In the thermodynamic limit this entropy diverges
at criticality.}\label{fig:fig1}
\end{figure}

Therefore we first consider the case of measuring just $J_{\pm k}$ and $I_{\pm k}$.  Using the results above, Eq.~\eqref{eq:Char} reduces to
\begin{align}
&\chi_{F}(u) = \\
&\frac{1}{Z_{k}(\lambda_0)Z_{-k}(\lambda_0)}\Tr({e}^{iu(\tilde{\mu}_k J_k +
\tilde{\mu}_{-k}J_{-k})}{e}^{-(1+iu)(\mu_k I_k +
\mu_{-k}I_{-k})}). \nonumber
\end{align}
Writing $\mathcal{N}_k(\lambda_0) = Z_k(\lambda_0)Z_{-k}(\lambda_0)$ we may
express the trace using the fermionic number states $\ket{n_k,n_{-k}}$ of the initial Hamiltonian
$H(\lambda_0)$ to find
\begin{align}
\chi_F(u) = &\frac{1}{\mathcal{N}_k(\lambda_0)}\sum_{n_{\pm k}=0,1}
\bra{n_k,n_{-k}}{e}^{iu(\tilde{\mu}_k J_k +
\tilde{\mu}_{-k}J_{-k})} \\
&\times{e}^{-(1+iu)(\mu_k I_k +
\mu_{-k}I_{-k})} \ket{n_k,n_{-k}}. \nonumber
\end{align}
These matrix elements are computable analytically and the analytic form of
the forward characteristic function is
\begin{align}
&\chi_F(u) =
\frac{1}{\mathcal{N}_k(\lambda_0)}\Bigl[{e}^{-2(1+iu)\mu_k}(e^{2iu\tilde{\mu}_k}\cos^{2}\Delta\alpha_k
+\sin^2\Delta \alpha_k)\nonumber \\
&+ 2{e}^{-(1+iu)\mu_k}{e}^{iu\tilde{\mu}_k} +(\cos^{2}\Delta\alpha_k
+{e}^{2iu\tilde{\mu}_k}\sin^2\Delta \alpha_k) \Bigr], 
\end{align}
where $\Delta \alpha_k =
\tilde{\alpha}_k-\alpha_k$ and we have used the fact $\mu_{k} = \mu_{-k}$ and $\tilde{\mu}_{k} = \tilde{\mu}_{-k}$.
Similarly, the backward characteristic function may be evaluated
straightforwardly:
\begin{align}
&\chi_B(v) =
\frac{1}{\mathcal{N}_k(\lambda_\tau)}\Bigl[{e}^{-2(1+iv)\tilde{\mu}_k}(e^{2iv\mu_k}\cos^{2}\Delta\alpha_k
+\sin^2\Delta \alpha_k)\nonumber \\
&+ 2{e}^{-(1+iv)\tilde{\mu}_k}{e}^{iv\mu_k} +(\cos^{2}\Delta\alpha_k
+{e}^{2iv\mu_k}\sin^2\Delta \alpha_k) \Bigr].
\end{align}
Setting $v = -u + i$ we find the ratio of the
forward and backward characteristic functions equals
$\mathcal{N}_k(\lambda_\tau)/\mathcal{N}_k(\lambda_0) =
Z_k(\lambda_\tau)Z_{-k}(\lambda_\tau)/Z_k(\lambda_0)Z_{-k}(\lambda_0)$
which is expected from the form of Eq.~\eqref{eq:eq5}.
Evaluating the derivative $\partial_{iu}\chi_F(u)|_{u=0} = \langle W \rangle$ we find the irreversible
work done under this nonequilibrium protocol is given by
\begin{align}
\langle W
\rangle &= \Bigl[2\tilde{\mu}_{k}e^{-2\mu_k}\cos^{2}\Delta\alpha_k -
2\mu_k{e}^{-2\mu_k}+2\tilde{\mu_k}{e}^{-\mu_k} \nonumber \\
&-2\mu_k{e}^{-\mu_k}+2\tilde{\mu}_k\sin^2\Delta\alpha_k \Bigr],
\end{align}
and hence the irreversible entropy is given by
\begin{equation}
\langle \Delta S^{irr} \rangle =
-\log\frac{\mathcal{N}_k(\lambda_\tau)}{\mathcal{N}_k(\lambda_0)} + \langle W
\rangle.
\end{equation}
 This quantity is plotted in Fig.~\ref{fig:fig1} as a function of the
pre-quench field $\lambda_0$ for different $k$ modes
and system size $N$ with fixed $\lambda_{t\ll 0} = 0$.  Focussing on the irreversible entropy associated
with measuring on the $k$ mode with the smallest magnitude $k = 2\pi/N$, we find a peak around  the quantum critical point $\lambda_0 = 1$. This peak sharpens with increasing $N$ and becomes singular in the thermodynamic
limit.  The intrepretation of this is simple:  if we prepare the system in a GGE
at the quantum critical point via a quench numerous excitations are produced due to vanishing energy gap.  
The existence of these excitations leads to the production of irreversible entropy on driving the system away from criticality.

Another interesting regime is where the initial and final Lagrange multipliers
are equal, but $\lambda_0 \neq \lambda_\tau$.  In the TFIM these conditions, combined with the spectral properties of $\{I_k\}$ and $\{J_k\}$, imply the ratio 
of the forwards- and backwards-protocol characteristic functions is unity and the irreversible entropy is always zero,
irrespective of any singular features of the model. This is in stark contrast to the case of a thermal state whereupon approaching the quantum
critical point the irreversible entropy in the thermodynamic limit diverges due
to the closing of the gap in the spectrum~\cite{Sachdev2011,Dorner2012}.  

Further evidence of the quantum critical point is provided from a general two
point measurement:  we now consider measuring the $I_k = \gamma^{\dagger}_k\gamma_{k}$ followed by a measurement of another mode's occupation $J_{k'} = \tilde{\gamma}^{\dagger}_{k'}\tilde{\gamma}_{k'}$, where $k' \neq k$.  Using the two-point measurement scheme the joint probability density [see Eq.~\eqref{eq:eq8}] and characteristic function are accessible
analytically.  From this we find the ``generalised work done'' during such a
protocol is 
\begin{align}
\label{eq:WTP}
\langle W^{TP} \rangle &= 
\Tr(\tilde{\gamma}^{\dagger}_{k'}\tilde{\gamma}_{k'}\rho_{GGE}(\lambda_0))-\Tr(\gamma^{\dagger}_k\gamma_k\rho_{GGE}(\lambda_0))
\nonumber
\\
&=\frac{1}{1+{e}^{\mu_k}}+\frac{1}{\mathcal{N}_{k'}}({e}^{-2\mu_{k'}}\cos^2\Delta
\alpha_{k'}+e^{-\mu_{k'}}\nonumber \\
&+\sin^2\Delta\alpha_{k'}).
\end{align}

\begin{figure}
\includegraphics[scale = 1.4]{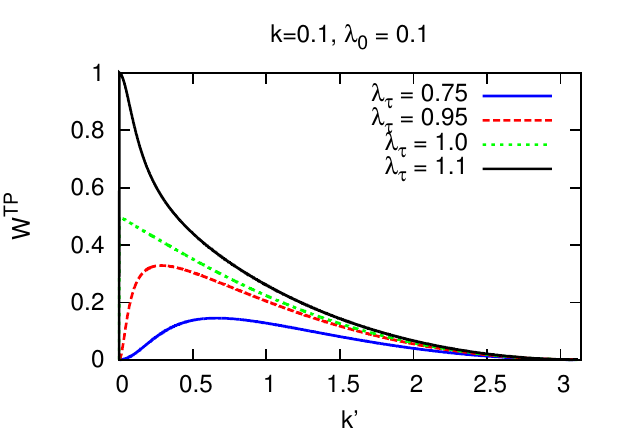}
\caption{(Color Online)  The work done as defined by the two point measurement
on modes $k = 0.1$ and $k'$ as a function of $k'$.  The initial GGE is prepared
via quench in the transverse field from $0\rightarrow 0.1$, while the GGE
compared with at the end of the protocol is prepared via quenches from
$0\rightarrow\lambda_\tau$.  When the final GGE is in the ferromagnetic phase
the protocol simply adds excitations to each $k$ mode.   The work done as a function $k'$ changes dramatically upon crossing the critical point with the occupation at $k' = 0$ growing the most compared to the other fermionic modes. }\label{fig:fig2}
\end{figure}

Fixing $\lambda_0 = 0.1$, $\lambda_{t\ll 0} = 0$ and $k = 0.1$, we plot the work
done as a function of $k'$ in Fig.~\ref{fig:fig2}, for various $\lambda_\tau$.  Effectively the
``work done'' is the difference in occupation of the fermionic modes.  The nonequilibrium protocol adds excitations to the system leading to the difference in occupation being positive, and increases as the protocol approaches the critical point $\lambda_\tau \rightarrow 1$.  At the critical point, the difference in occupation at $k' = 0$ equals 1/2.  If the nonequilibrium protocol crosses the critical point many excitations are produced and this leads the occupation difference to approach unity as $k' \rightarrow 0$. The fact that the largest change in occupation occurs at $k' = 0$ is consistent with the gap closing around this wavevector at criticality.

\emph{Summary.}  We have derived appropriate fluctuation theorems for integrable systems
whose equilibrium states are described be GGEs.  We take the archetypal
example of the TFIM and apply non-equilibrium protocols to different
modes.  We find a peak in the irreversible entropy when quenching around
the critical point which sharpens as the wavelength of the mode
decreases and system size increases.  This we can understand in terms of
the closing of a gap at criticality.  Correspondingly the closing of the gap manifests itself within a simpler two-point scheme on different modes and quench across the critical point.  Our results
will be of significance for nearly-integrable systems which exhibit
prethermalisation:  our protocols suggest out-of-equilibrium tests for
GGEs, the results of which depend on the timescale involved in
preparing the initial GGE state where it may only be metastable~\cite{Matteo2013}.  


\begin{acknowledgements}
We thank Matteo Marcuzzi for fruitful discussions and
comments on the manuscript. This work was supported by the European Union
Research Scholarship provided by the University of Nottingham and the Leverhulme Trust grant no.~F/00114/BG.
\end{acknowledgements}

%

\end{document}